\documentclass[aps,prl,twocolumn,epsfig]{revtex4}
\usepackage{amssymb}

\usepackage[dvips]{graphicx}
\usepackage{dcolumn}
\usepackage{bm}

\input{tcilatex}

\begin{document}

\title{Vortices near surfaces of Bose-Einstein condensates}
\author{J.R. Anglin \\
Center for Ultracold Atoms, MIT 26-251, 77 Massachusetts Ave., Cambridge MA
02139}

\begin{abstract}
The theory of vortex motion in a dilute superfluid of inhomogeneous density
demands a boundary layer approach, in which different approximation schemes
are employed close to and far from the vortex, and their results matched
smoothly together. \ The most difficult part of this procedure is the
hydrodynamic problem of the velocity field many healing lengths away from
the vortex core. \ This paper derives and exploits an exact solution of this
problem in the two-dimensional case of a linear trapping potential, which is
an idealization of the surface region of a large condensate. \ It thereby
shows that vortices in inhomogeneous clouds are effectively `dressed' by a
non-trivial distortion of their flow fields; that image vortices are not
relevant to Thomas-Fermi surfaces; and that for condensates large compared
to their surface depths, the energetic barrier to vortex penetration
disappears at the Landau critical velocity for surface modes.
\end{abstract}

\maketitle

\affiliation{Center for Ultracold Atoms, \\
Massachusetts Institute of Technology 26-251, \\
77 Massachusetts Avenue, Cambridge 02139} 


\section{Introduction}

Although much has long been known about quantized vortices in superfluids of
constant homogeneous density, vortices in trapped dilute Bose-Einstein
condensates move and interact within significantly inhomogeneous background
clouds. \ The motion of vortices in an inhomogeneous superfluid, including
their stability and location in a rotating condensate, has been the subject
of several recent theoretical works, both numerical \cite
{FCS,JMA,Anomalous,AD1,TKU,FC}\ and analytic \cite
{RP,Pismen,CastinDum,PeterGora,SF1,GG,SF2,KPSZ}. \ After years of effort,
vortices in dilute condensates are now also a subject of active experimental
study \cite{JILA1,JILA2,ENS1,ENS2,MIT1,MIT2,JILA3,Oxford}. \ \ Quantitative
comparison between theory and experiment therefore requires refinement of
theoretical methods, in order to go beyond qualitative or order-of-magnitude
estimates, and derive more rigorous predictions. \ 

This paper will take a step in this direction, by providing an exact
solution (in a non-trivial but two-dimensional case) to a part of the
problem that has usually been treated with uncontrolled approximations. \
Our specific conclusions will include the verdicts that image vortices are
inapplicable to surface effects in experimental condensates, and that the
often-cited energetic barrier to vortex penetration \cite{Anomalous}\
disappears at velocities no greater than the surface mode critical velocity 
\cite{FedMur,me}, well below previous estimates \cite{SF2,FCS}. \ This
conclusion applies in the experimentally relevant limit of a condensate
large compared to its surface depth, and the lowering of critical rotation
frequencies which it implies is an addition to any effects of asymmetric
potentials \cite{KPSZ}. \ The case we examine also provides a general
warning against neglecting the nontrivial hydrodynamics of inhomogeneous
superfluids.

\subsection{Boundary layer theory \ }

\ In a dilute superfluid whose macroscopic wavefunction $\Psi $\ is governed
by the Gross-Pitaevskii equation 
\begin{equation}
i\hbar \partial _{t}\Psi =-\frac{\hbar ^{2}}{2M}\nabla ^{2}\Psi +V(\vec{x}%
)\Psi +\frac{\hbar ^{2}g}{M}|\Psi |^{2}\Psi \;,  \label{GPE}
\end{equation}
the core diameter of a vortex at position $\vec{x}_{0}$ is on the order of
the local \textit{healing length }$\xi =\left[ g\rho \left( \vec{x}%
_{0}\right) \right] ^{-1/2}$, where $M$ is the particle mass, $g$ is the
interaction strength, and $\rho \equiv |\Psi |^{2}$ is the condensate number
density, which varies smoothly over the sample from zero to its maximum
value. \ This maximum is fixed by the chemical potential $\mu $, which is
not a parameter in the Gross-Pitaevskii equation, but an integration
constant: the general solution admits a time-dependent prefactor in $\Psi $
of the form $e^{-i\mu t/\hbar }$. \ 

Throughout this paper we will consider a quasi-two-dimensional condensate,
so that $g$ is dimensionless \cite{2Dg}. \ Our results may also be applied,
with trivial changes, to the case of a hydrodynamically 3D condensate in
which the density is constant along the third direction, and the vortex
lines are parallel to it. \ In three-dimensional background clouds, however,
vortex lines generally bend \cite{SF2,Bending,Pismen}, and this introduces
complications which usually prohibit accurate analytical treatment \cite
{Pismen}; \ hence our restriction, in this paper, to the two-dimensional
case. \ Even in the limit of a quasi-two-dimensional condensate, however,
the inhomogeneous background density of a trapped condensate significantly
affects the behaviour of two-dimensional `point vortices'. \ The
Thomas-Fermi (TF) surface, which is the edge of the condensate cloud in the
hydrodynamic approximation \cite{LPS}, also affects vortex motion.

It is typically the case, in current experiments, that near the vortex the
trapping potential $V$ varies slowly on the healing length scale. \ This
small ratio of scales has two distinct implications for vortex physics. \
Firstly, it means that within many healing lengths around the vortex center,
the gradient in $V$ can be treated as a perturbation. \ Secondly, it means
that beyond a few healing lengths from the vortex center, we may solve (\ref
{GPE}) in the \textit{hydrodynamic approximation}. \ These are two very
different kinds of approximation, but neither is valid everywhere, and so
they must be combined. \ Fortunately there is a significant intermediate
region near the vortex within which both approximations are valid, and this
will allow us to match the solutions they yield smoothly together, to obtain
a complete solution. \ This is an example of a general procedure, known as
boundary layer theory, or the method of matched asymptotics. \ 

Boundary layer theory is more than just a technical trick without physical
meaning: it is an essential feature of their physics that vortices are
small, healing-length-scale structures, realized within a larger-scale
hydrodynamic medium. \ \ Boundary layer theory is a very direct expression
of the multiple-scale nature of vortices, and the only alternatives to it
are exact (or numerical) solutions, which will of course also exhibit the
problem's multiple scales, or the replacement of some part of the boundary
layer analysis with some uncontrolled approximation. \ The boundary layer
method has yielded general equations of motion for dark solitons in
quasi-one-dimensional inhomogeneous backgrounds \cite{BA}, and has already
been applied to 2D vortex motion in inhomogeneous backgrounds \cite{RP,SF2}.
\ In this paper we will take a precisely similar approach. \ Our main
contribution is an exact solution to the outer, hydrodynamic part of the
problem, in a physically relevant and nontrivial case. \ Comparing our exact
solution to the assumptions\textit{\ }of previous work will illustrate some
general aspects of vortex physics that may not have been fully appreciated
heretofore. \ 

\subsection{Organization}

This paper is organized as follows. \ In the following Section II we review
the hydrodynamic approximation, and then identify a loophole in previous
derivations of a general solution supposed to be valid in the neighbourhood
of a vortex. \ We show that in fact a nontrivial problem must be solved,
which generically lacks small parameters and involves the whole condensate
globally. \ We then present a physically relevant case in which this problem
can be solved exactly, in terms of a special function which is only
moderately obscure, and whose asymptotic behaviours close to and far from
the vortex can be obtained analytically. \ 

In Section III we follow earlier authors \cite{RP}\ in obtaining the inner
solution, perturbing around the numerical solution to the Gross-Pitaevskii
equation for a vortex in constant background density. \ We then go on to
match the inner and outer solutions together, and thereby determine the
velocity at which the vortex moves parallel to the TF surface. \ With these
results, we compute the free energy of a vortex in a moving frame, and thus
assess the velocity at which the energetic barrier preventing vortices from
entering the condensate will disappear. \ We compare these results to recent
analogous calculations for critical rotation frequencies of harmonically
trapped condensates, concluding that the latter overestimate critical
frequencies of large condensates by factors of order unity. \ 

In our final Section IV we discuss our results, interpreting the vortex
motion as due to vortex buoancy, through a Magnus effect which is
renormalized by the distortion of the flow field. \ We interpret this
distortion as an `infrared dressing' of the vortex, and emphasize that such
dressing is a much more general effect than can be described with image
vortices. \ We argue that the energetic barrier to vortices actually
disappears at or below the surface mode critical velocity \cite{FedMur,me},
so that, strictly speaking, vortices entering condensates should not be said
to `nucleate' (cross an energy barrier by thermal fluctuations or quantum
tunneling). \ We then conclude with a brief summary and outlook. \ 

\section{Outer solution: hydrodynamics}

\subsection{Hydrodynamic approximation}

The hydrodynamic approximation works as follows. \ If we define $\Psi =\sqrt{%
\rho }e^{i\theta }$ for real $\rho ,\theta $, the Gross-Pitaevskii equation
becomes exactly 
\begin{eqnarray}
\partial _{t}\rho  &=&-\frac{\hbar }{M}\vec{\nabla}\cdot \rho \vec{\nabla}%
\theta \text{ \ \ (continuity)}  \label{continuity} \\
\partial _{t}\theta  &=&-\frac{\hbar }{M}\left[ \frac{1}{2}|\vec{\nabla}%
\theta |^{2}+\frac{M}{\hbar ^{2}}V+\frac{1}{\xi ^{2}}\frac{\rho }{\rho _{0}}-%
\frac{\nabla ^{2}\sqrt{\rho }}{2\sqrt{\rho }}\right] \;  \label{TF} \\
\rho _{0} &\equiv &\rho \left( \vec{x}_{0}\right) ,  \nonumber
\end{eqnarray}
where the vortex center is located at $\vec{x}_{0}$. \ We write $g$ in the
form $1/\left( \xi ^{2}\rho _{0}\right) $ in order to introduce the vortex
scale $\xi $ explicitly. \ We seek a solution in which the only time
dependence is due to the vortex motion, so that sufficiently far from $\vec{x%
}_{0}$ we can set $\dot{\rho}=0$, and $\dot{\theta}=-\mu /\hbar $ with
constant $\mu $. \ If $\theta $ and $V$ are functions of $\varepsilon \vec{x}%
/\xi $ for some small $\varepsilon ,$ then the second equation (\ref{TF})
just yields the Thomas-Fermi density profile $\rho _{TF}(\vec{x})=M\left(
\mu -V\right) /\left( \hbar ^{2}g\right) $, up to corrections of order $%
\varepsilon $ or higher. \ The surface on which $\rho _{TF}$ vanishes is the
TF surface, where the TF approximation breaks down. (An additional boundary
layer treatment is therefore required very close to the TF surface \cite{LPS}%
.)

In this case we may expect (and later confirm) that the vortex velocity $%
v_{vtx}$ is order $\varepsilon c$, where $c=\hbar /\left( M\xi \right) $ is
the speed of bulk sound near the vortex. This means that the corrections are
formally of order $\varepsilon ^{2}$. \ But there will also be corrections
that diverge as one approaches the vortex, the leading one being the vortex
kinetic energy $\propto r^{-2}$. \ To keep these corrections no larger than
order $\varepsilon $, we can only use the hydrodynamic approximation in an
`outer zone' whose inner boundary is a circle around the vortex centre,
having radius $R$ of order $\varepsilon ^{-1/2}\xi $. \ In the `inner zone'
inside this radius, we will use perturbation theory on the potential
gradient instead. \ As long as it has the right order of magnitude, the
precise value of $R$ is arbitrary: it is merely a bookkeeping device to let
us merge two different approximations, and all $R$-dependent terms
necessarily cancel when the two zones are patched together. \ So the full
calculation will in fact be accurate up to corrections of order $\varepsilon
^{2}$ after all. \ And this definition of the outer zone also formalizes the
requirement of being `sufficiently far' from the vortex to set $\dot{\rho}=0$
and $\dot{\theta}=-\mu /\hbar $, since for $\left| \vec{x}-\vec{x}%
_{0}\right| >R$ the corrections $\vec{v}_{vtx}\cdot \vec{\nabla}\rho ,\theta 
$ will be of order $\varepsilon ^{2}$. \ 

The higher order corrections can all be computed trivially once the phase
field $\theta $ is known to zeroth order. \ (We will not actually compute
such corrections in this paper.) \ The zeroth order phase field can be
obtained by solving the continuity equation (\ref{continuity}) with the
zeroth order TF density, and so this is the main problem of the outer zone.
\ 

\subsubsection{Dual fields}

Because the phase will be multi-valued in the presence of vortices, it is
convenient in two dimensions to define the field $\vec{A}$ which is dual to
the velocity field (in the sense in which electric and magnetic fields may
be dual): 
\begin{equation}
(A_{x},A_{y})\equiv \left( \partial _{y}\theta ,\;-\partial _{x}\theta
\right) \;.  \label{dualA}
\end{equation}
The continuity equation then becomes $\vec{\nabla}\times (\rho \vec{A})=0$,
which can be identically satisfied by setting 
\begin{equation}
\vec{A}\equiv \frac{\rho _{0}}{\rho }\vec{\nabla}F  \label{Fdef}
\end{equation}
for some potential $F(x,y)$. \ \ We will refer to $F$ as the dual potential
to the phase $\theta $, but it is also known as the streamline function,
because\ its contours of constant height are streamlines of the velocity
field. \ Because it is single-valued even when vortices are present, $F$ is
a more convenient substitute for $\theta $.\ \ \ $F$ is not arbitrary,
because we can see from (\ref{dualA}) that $\vec{\nabla}\cdot \vec{A}=\vec{%
\nabla}\times \vec{\nabla}\theta =2\pi \,\delta ^{2}(\vec{x}-\vec{x}_{0})\,($%
since a singly quantized vortex is located at $\vec{x}_{0}$). \ Hence $F$
must satisfy $\rho _{0}\vec{\nabla}\cdot \left( \rho ^{-1}\vec{\nabla}%
F\right) =2\pi \,\delta ^{2}(\vec{x}-\vec{x}_{0}).\,$ \  But since we are
only seeking the zeroth order phase field, we can replace $\rho \rightarrow
\rho _{TF}$ in both (\ref{Fdef}) and in the constraint on $F$, writing $\vec{%
A}\doteq \left( \rho _{0}/\rho _{TF}\right) \vec{\nabla}F$ as well as 
\begin{equation}
\rho _{0}\vec{\nabla}\cdot \frac{\vec{\nabla}F}{\rho _{TF}}\doteq 2\pi
\,\delta ^{2}(\vec{x}-\vec{x}_{0})\,  \label{dualsource}
\end{equation}
where the $\doteq $ means the neglect of terms that will ultimately lead to
corrections of order $\varepsilon ^{2}$. \ (Introducing $_{0}$ subscripts on 
$F$ and $\vec{A}$ would too greatly encumber our notation.) \ Eqn. (\ref
{dualsource}) is the essential equation determining the zeroth order outer
zone phase field of a vortex. \ Note that since it is a linear equation,
solutions with more vortices can be obtained trivially from the single
vortex case. \ 

As a simple example, consider the classic case of a vortex in a constant
background density profile, near a hard wall. \ Choose the $y$ axis to run
along the wall, and let the vortex be located on the $x$ axis at the point $%
x_{0}$. \ Since in this case $\rho /\rho _{0}=1$, we merely have a Laplace
equation to solve, with a delta-function source, and the boundary condition
that there be no flow through the line $x=0$. \ Since adding a constant to $F
$ obviously does nothing, we can enforce this boundary condition by setting $%
F(0,y)=0$. \ This is satisfied by the solution 
\[
F_{HW}=\frac{1}{2}\ln \frac{(x-x_{0})^{2}+y^{2}}{(x+x_{0})^{2}+y^{2}}
\]
where a possible extra term proportional to $x$ must vanish in the case that
the velocity field vanishes at infinity. \ This solution can evidently be
obtained by the method of images, with the denominator of the logarithm
representing the dual potential of an image antivortex located at $-x_{0}$.
\ 

Obtaining an analogous solution for an inhomogeneous $\rho $ is generally
much more difficult, however.\ \ 

\subsubsection{Proposed general results}

General equations of motion for point vortices in inhomogeneous superfluid
backgrounds have nevertheless been presented, having been derived using the
boundary layer approach \ \cite{RP,SF1}. \ According to these proposed
equations, vortex motion is determined by the local trapping potential in
the immediate neighbourhood of the vortex (that is, can be given by a
universal formula involving the potential, its gradient, and its Laplacian
at $\vec{x}_{0}$). \ We pause here to point out a loophole in the derivation
of these results, which unfortunately allows corrections to vortex motion
that are comparable in size to the proposed general results, and that in
general cannot be computed without solving the global hydrodynamic problem.
\ \ 

The approach of Rubinstein and Pismen, and of Svidzinsky and Fetter
following them, is the same approach we are following, of matching a
hydrodynamic outer zone with a perturbative inner zone. \ At the point we
have now reached, namely solving Eqn. (\ref{dualsource}), these authors
employ the variable $H(x,y)=\sqrt{\rho _{0}/\rho }F(x,y)$ (Ref. \cite{SF1}
uses a more elaborate notation for the same function) and so obtain the
equivalent equation, 
\begin{eqnarray}
\left[ \nabla ^{2}-k^{2}\right] H &=&2\pi \,\delta (x-x_{0})\,\delta (y)
\label{RP1} \\
\text{for \ }k^{2} &\equiv &\frac{1}{2}\left[ -\frac{\nabla ^{2}\rho }{\rho }%
+\frac{3}{2}\left| \frac{\vec{\nabla}\rho }{\rho }\right| ^{2}\right] . 
\nonumber
\end{eqnarray}
\ Their general procedure, for any slowly varying potential, is then to
replace $k(x)\rightarrow k(x_{0})=k_{0}$, so that (\ref{RP1}) becomes simply
a 2D Helmholtz equation (with a delta-function source). \ \ It is then
observed that the modified Bessel function $H(x,y)=-K_{0}\left( k_{0}\sqrt{%
(x-x_{0})^{2}+\left( y-y_{0}\right) ^{2}}\right) $ is a particular solution
to (\ref{RP1}) with $k(x)\rightarrow k_{0}$. \ Of course only the $%
(x,y)\rightarrow (x_{0},y_{0})$ limit of this result is to be taken
seriously, so the Rubinstein-Pismen result for the outer solution near the
vortex is 
\begin{eqnarray*}
&&F_{RP}\left( x_{0}+\xi r\cos \phi \;,\;y_{0}+\xi r\sin \phi \right)  \\
&=&\sqrt{1+\xi \rho _{0}^{-1}\vec{r}\cdot \vec{\nabla}\rho _{0}}\left(
\gamma +\ln \frac{r\xi k_{0}}{2}\right) +\mathcal{O}(\varepsilon ^{2}).
\end{eqnarray*}
Euler's constant $\gamma \doteq 0.577$ appears through the asymptotic
behaviour of the modified Bessel function at small argument. \ 

The problem with this computation is that it assumes that the solution for $%
H $\ contains no components regular at $(x,y)\rightarrow (x_{0},y_{0})$
(terms like $I_{0},I_{1}\cos \phi $, etc.). \ Actually, nothing ensures
this. \ And so all that can really be claimed, on the basis of purely local
analysis near the vortex, is that 
\begin{eqnarray}
\lim_{\vec{x}\rightarrow \vec{x}_{0}}F &=&\left( 1+\frac{\xi }{2}\vec{r}%
\cdot \vec{\nabla}\rho _{0}\right) \ln r  \nonumber \\
&&+k_{0}\xi r(B_{y}\cos \phi -B_{x}\sin \phi )+\mathcal{O}(\varepsilon ^{2})
\label{F1a}
\end{eqnarray}
for some undetermined constants $B_{x,y}$. \ (A constant term can be dropped
because it will not contribute to the dual velocity field $\vec{A}$.) \ \
This unknown $\vec{B}$ implies an unknown correction to the superfluid
velocity field near the vortex, and hence to the vortex velocity. So the
strictly local approach, based on solving (\ref{RP1}) with $k(\vec{x}%
)\rightarrow k_{0}$, does not actually lead to any conclusion at all for the
vortex motion.\ 

In Reference \cite{RP}, Rubinstein and Pismen are careful to state that they
are deriving the vortex velocity relative to the `ambient' superfluid
velocity, thus recognizing the corrections they are omitting. \ It is worth
emphasizing, however, that the fluid velocity that may be called `ambient'
is in a sense scale dependent. \ The effect omitted in Refs. \cite{RP,SF1}
is indeed a flow near the vortex that is constant \textit{on the healing
length scale}; but it is generally \textit{not} constant on the much longer
hydrodynamic scale of the outer zone. \ So while such a flow may be called
`ambient' from the point of view of the vortex core, in general it is 
\textit{not} simply the flow that is specified at infinity (or wherever
boundary conditions are imposed).\ \ 

Whereas Rubinstein and Pismen recognize but omit them, Svidzinsky and Fetter
argue in Reference \cite{SF1} that additional terms regular at the vortex
may be ruled out, because the $I_{n}$ functions all diverge at large
argument, and so would violate any physical boundary conditions. \ While it
is true that the $I_{n}$ all diverge at large argument, they only do this on
the scale $k_{0}$, and once $k_{0}|\vec{x}-\vec{x}_{0}|$ is no longer small,
the approximation $k(\vec{x})\rightarrow k_{0}$ is no longer good. \ Hence
the argument\ of \cite{SF1} is invalid.

\subsubsection{Motivation for an exact solution}

The moral of this general discussion should be clear. \ Implementing the
hydrodynamic approximation has exhausted the benefits of the slow variation
of the trapping potential on the healing length scale and thus, in general,
Eqn. (\ref{dualsource}) contains no small parameters.\ So except in special
cases where additional small parameters may appear, such as for a trap with
a high aspect ratio, or a vortex very close to the center of symmetric trap,
no accurate equation of motion for a vortex may be obtained without solving
a nontrivial hydrodynamic problem exactly. \ 

In\ \cite{RP}, Rubinstein and Pismen provide one nontrivial exact solution,
for the case described in our variables as $M[\mu -V(\vec{x})]=$ $\hbar
^{2}g\rho _{0}[I_{0}(r/\mathcal{R})]^{-2}$ for some $\mathcal{R}\gg \xi $. \
They focus on this case because it ensures that $k(x)$ in (\ref{RP1}) really
is constant, so that their solution quoted above becomes exact. \ The trap
required to realize their solvable example with trapped dilute condensates
is not implausible (it closely resembles a Gaussian well); but one would
also need a finely tuned chemical potential, to make the condensate density
just vanish at infinity. \ And it is a special feature of this finely-tuned
case that there is no Thomas-Fermi radius. \ (Corrections to the
Thomas-Fermi profile begin to exceed $\left( \xi /\mathcal{R}\right) ^{2}$
for $r>\mathcal{R}\ln \left( \mathcal{R}/\xi \right) $, but their onset is
very gradual.) \ It is therefore the main contribution of this paper to
provide, in the next subsection, another instructive exact solution to the
hydrodynamic problem of a superfluid vortex in an inhomogeneous background,
which is more directly relevant to currently typical experiments.

\subsection{Vortex hydrodynamics in a linear density profile}

\subsubsection{The plane linear approximation}

Near a Thomas-Fermi surface, the potential is approximately linear; and the
TF surface of a large condensate is approximately flat. \ This motivates
considering the idealized problem of a linear ramp potential \cite
{LPS,AKPS,FedMur,me}. \ Choosing the $y$ axis to run along the straight TF
surface, we have the TF profile 
\begin{mathletters}
\begin{equation}
\rho =\rho _{0}\frac{x}{x_{0}}\text{.}  \label{TFlin}
\end{equation}
Since it will turn out that $F$ decays on the distance scale of $x_{0}$, the
linear ramp potential will indeed be reasonably accurate for real traps,
which are obviously not globally linear, as long as the vortices are not too
far from the TF surface; for harmonic traps, this means that $x_{0}$ should
be much less than the TF radius. \ On the other hand, with $V=\mu -\lambda x$
(obtained automatically by taking $x=0$ on the TF surface)$,$ we have $\xi
=\hbar /\sqrt{M\lambda x_{0}}$, and hence $\varepsilon =\xi /x_{0}=\hbar /%
\sqrt{M\lambda x_{0}^{3}}$. \ The hydrodynamic approximation breaks down
within a few surface depths of the TF surface, where the surface depth \cite
{LPS} is 
\end{mathletters}
\begin{equation}
\delta =\left( \hbar ^{2}/2M\lambda \right) ^{1/3}.
\end{equation}
We therefore also have $\varepsilon =\sqrt{2}\left( \delta /x_{0}\right)
^{3/2}$, and so our boundary layer treatment of the vortex will require $%
x_{0}\gg \delta $. \ See Figure 1 for a sketch of our model system
indicating the relationships between the various length scales that will
appear in our calculations. \ 

Vortices located well outside the TF surface ($x_{0}\ll -\delta $) can be
described perturbatively as surface excitations of the condensate \cite{me};
but vortices with centers within the surface layer $|x|\lesssim \delta $
require a nonperturbative treatment, which we are unable to provide. \ In
our final Section, though, we will argue that nothing remarkable can happen
in this regime, which is analytically intractable but physically trivial. \
As mentioned above, the TF surface requires a boundary layer treatment of
its own \cite{LPS}, independent of the vortex; but the post-hydrodynamic
effects within this layer can easily be shown to produce corrections smaller
than all results we will report by factors of at least $(\delta /x_{0})^{2}.$
\ Hence for vortices many surface depths into the condensate, we will be
able to ignore completely the breakdown of hydrodynamics at the TF surface
itself. \ 

\begin{figure}[tbp]
\includegraphics[width=0.95\linewidth]{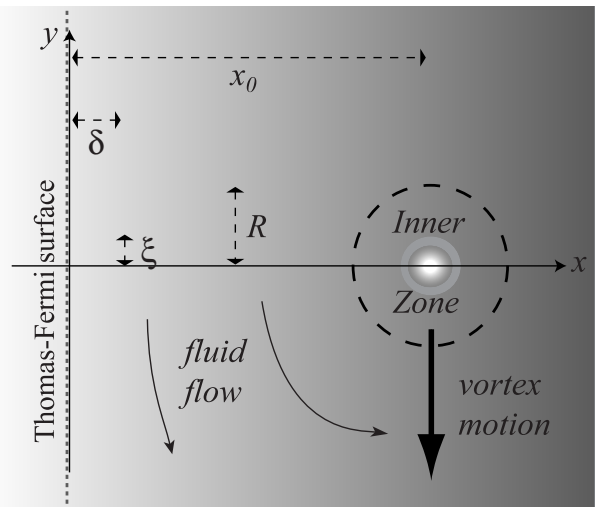}
\caption{Sketch with scales. \ Darkness of the background indicates
condensate density, increasing linearly with distance $x$ from the
Thomas-Fermi surface at left, on which the $y$ axis is placed. \ The
hydrodynamic approximation is used in the Outer Zone, further than $R$ from
the vortex center; in the Inner Zone the full Gross-Pitaevskii equation is
solved with the potential gradient treated as a perturbation. \ The relative
sizes of the four scales $\protect\xi ,\protect\delta ,R$ and $x_{0}$ are
indicated qualitatively, but the ratios should obviously be rather greater
for the analysis in the text to be accurate. \ The counter-clockwise
circulating vortex is shown in the text to move in the negative $y$
direction.}
\end{figure}

\subsubsection{The equation}

Setting $y_{0}=0$ by our choice of origin,\ Eqn. (\ref{dualsource}) in this
case becomes 
\begin{equation}
\lbrack \partial _{xx}-\frac{1}{x}\partial _{x}+\partial _{yy}]F=2\pi \delta
(x-x_{0})\,\delta (y).  \label{hydroedge}
\end{equation}
\ As warned above, this equation has no small parameters in it: even $x_{0}$
may be scaled away. \ Fortunately, however, (\ref{hydroedge}) is exactly
solvable. \ Taking advantage of translational symmetry in $y$, we can write\
\ \ \ 
\begin{equation}
F(x,y)=\int_{-\infty }^{\infty }dk\,f_{k}(x)\,e^{iky}
\end{equation}
to obtain the ODE 
\begin{equation}
f_{k}^{\prime \prime }-\frac{1}{x}f_{k}^{\prime }-k^{2}f_{k}=\delta
(x-x_{0})\,\;.  \label{disc}
\end{equation}
We begin by solving the homogeneous equation, setting the RHS to zero. \
Multiplying $f_{k}$ by $x$ allows us to recognize a modified Bessel
equation, and so we obtain solutions involving modified Bessel functions of
order 1: 
\begin{equation}
f_{k}^{0}=x\left[ A_{k}K_{1}(kx)+B_{k}I_{1}(kx)\right] 
\end{equation}
for constants $A_{k},B_{k}$. \ We wish to impose the hydrodynamic boundary
condition of no velocity through the TF surface, corresponding to $F(0,y)=$
a constant that we can choose to be zero; so since $\lim_{x\rightarrow
0}K_{1}(x)=1/x$, we need $A_{k}=0$. \ We can accept no solutions that grow
exponentially as $x\rightarrow \infty $, and so we must also set all $B_{k}=0
$. \ The only exception that might occur is the special case $k=0$, in which
we might under some circumstances require $f_{0}^{0}=B_{0}x^{2}$, since $%
F\propto x^{2}$ corresponds merely to a constant velocity field in the $y$
direction.  But for the case where the velocity vanishes at infinity, we
need $B_{0}=0$ as well.

Note the difference here between our rejection of additional homogeneous
solutions, and their rejection in \cite{SF1}: we have solved our
hydrodynamic problem exactly for all $x$, and not just asymptotically near
the vortex. \ Of course, to apply our model to a realistic case we must
admit that the potential becomes nonlinear at sufficiently large $x$ or $y$;
but as long as this nonlinearity scale (call it $\Lambda $) is much larger
than $x_{0}$, our exact solution will be applicable over an essentially
infinite region, as far as the vortex is concerned. \ One loophole does
remain, however, even in this case. \ We have fixed $B_{0}=0$ by demanding
that the velocity field vanish at infinity, to which we assume that our
linear model of the potential extends. \ But in a real, finite system, even
if it is modelled well by a linear potential over a large region, it could
be that boundary conditions farther away from the vortex than $\Lambda $
imply some nontrivial, nonzero $B_{0}$. \ This would allow corrections to
the vortex velocity that one can expect to be of order $\hbar /(\Lambda M)$.
\ \ To compute them would require solving the actual, finite hydrodynamic
problem.

\subsubsection{The vortex solution}

Having solved our homogeneous equation in general and concluded that the
only solution admitted by our boundary conditions is zero, there remains the
main task of obtaining a particular solution for $F$\ with the delta
function source at $x=x_{0}.$ \ Since we have the general solutions to the
homogeneous equation, we can patch them together to meet our boundary
conditions and also fit the delta function source, by writing 
\begin{equation}
f_{k}=C\frac{x}{x_{0}}\times \left\{ 
\begin{array}{c}
K_{1}(kx_{0})I_{1}(kx)\;,\;x<x_{0} \\ 
I_{1}(kx_{0})K_{1}(kx)\;,\;x>x_{0}
\end{array}
\right\} 
\end{equation}
for a constant $C$. \ We then fix $C$ by imposing that the discontinuity in
the derivative of $f_{k}$\ at $x=x_{0}$ be equal to $1$, as required by (\ref
{disc}). \ Since $K_{1}^{\prime }(\xi )I_{1}(\xi )-K_{1}(\xi )I_{1}^{\prime
}(\xi )=-1/\xi $, this gives $C=-1$, and so we have

\begin{eqnarray}
F &=&-2\frac{x}{x_{0}}\int_{0}^{\infty }dk\,\cos
ky\,[K_{1}(kx_{0})I_{1}(kx)\theta (x_{0}-x)  \nonumber \\
&&\;\;\;\;\;\;\;\;\;\;\;+K_{1}(kx)I_{1}(kx_{0})\theta (x-x_{0})]\;
\end{eqnarray}
where $\theta (x)$ is the step function. \ Happily, this integral can be
evaluated in closed form in terms of a Legendre function of the second kind 
\cite{GunsRoses}, yielding 
\begin{equation}
F(x,y)=-\sqrt{x/x_{0}}Q_{\frac{1}{2}}\left( z\right) \;,  \label{Qsol}
\end{equation}
where $z=$ $z(\frac{x}{x_{0}},\frac{y}{x_{0}})$ is defined as 
\begin{equation}
z=\frac{x^{2}+y^{2}+x_{0}^{2}}{2xx_{0}}=1+\frac{(x-x_{0})^{2}+y^{2}}{2xx_{0}}
\label{zdef}
\end{equation}
The Legendre functions of order $\nu $ satisfy the Legendre differential
equation 
\begin{equation}
\frac{d}{dz}\left[ \left( 1-z^{2}\right) \frac{dQ_{\nu }}{dz}\right] =-\nu
\left( \nu +1\right) Q_{\nu }\;  \label{leg12}
\end{equation}
and the functions of the second kind have logarithmic singularities at $%
z=\pm 1$. \ \ It is straightforward to verify, using (\ref{leg12}), that (%
\ref{Qsol}) satisfies (\ref{dualsource}) with the boundary condition $%
F(0,y)=0$. \ 

The vortex flow field can be visualized by plotting contour lines of
constant $F$\ which correspond to fluid flow lines. \ Such contour plots are
available numerically through common commercial software applications that
include large libraries of special functions. \ An example is shown in
Figure 1, with the corresponding flow pattern for a vortex in homogeneous
density near a hard wall, for comparison. \ It is obvious that the density
gradient distorts the flow pattern signficantly. 
\begin{figure}[tbp]
\includegraphics[width=0.95\linewidth]{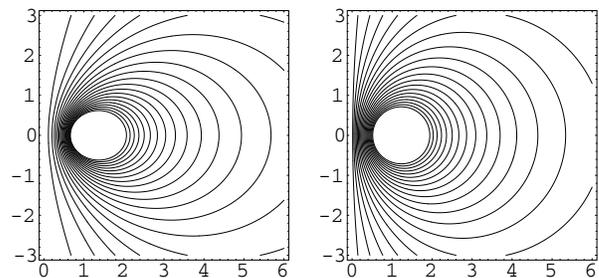}
\caption{Hydrodynamic flow lines around a 2D point vortex near (left) a TF
surface and (right) a hard wall. In both cases the surface of the condensate
is the vertical axis, with the axes marked in units of $x_{0}$. The left
plot is constant height contours of $F=-\protect\sqrt{x/x_{0}}Q_{1/2}(z)$
for $z(x,y)$ defined in the text. The right plot consists of constant height
contours of the homogeneous dual potential with a hard wall surface, $%
F_{HW}=\ln [\protect\sqrt{(x-x_{0})^{2}+y^{2}}/\protect\sqrt{%
(x+x_{0})^{2}+y^{2}}]$,.}
\end{figure}

\subsubsection{Outer asymptotics}

At large $z$, $Q_{1/2}\rightarrow 2^{-5/2}\pi z^{-3/2}$ (easily obtained
from (\ref{qintrep}) below, or see\ \cite{Guns2}), so that the dual
potential $F$ falls off at long range much more quickly than the long-range
logarithmic behaviour found at constant background density. \ Inserting our
definition of $z$\ we therefore have 
\begin{eqnarray}
\lim_{|y|\rightarrow \infty }F(x,y) &=&-\frac{\pi x^{2}x_{0}}{2|y|^{3}}%
\left( 1+\frac{x^{2}}{y^{2}}\right) ^{-3/2}  \label{as1} \\
\lim_{x\rightarrow \infty }F(x,y) &=&-\frac{\pi x_{0}}{2x}\left( 1+\frac{%
y^{2}}{x^{2}}\right) ^{-3/2}  \label{as2} \\
\lim_{x\rightarrow 0^{+}}F(x,y) &=&-\frac{\pi x^{2}}{2x_{0}^{2}}\left( 1+%
\frac{y^{2}}{x_{0}^{2}}\right) ^{-3/2}.  \label{as3}
\end{eqnarray}
The kinetic energy density due to the vortex thus falls off as $x^{-5}$ at
large $x$ for fixed $y$, and as $xy^{-6}$ for large $y$ and fixed $x$, which
allows us to conclude that vortices near Thomas-Fermi surfaces are
essentially localized structures that do not extend into the condensate
beyond a depth of order $x_{0}$.

\subsubsection{Inner asymptotics}

For matching with the inner zone solution, which will determine the vortex
velocity, we need the behaviour of $Q_{1/2}(z)$ as $z\rightarrow 1^{+}$. \
This can be obtained analytically. \ Using dimensionless polar co-ordinates $%
\left( r,\phi \right) $ centred on the vortex,

\begin{equation}
x=x_{0}+\xi r\cos \phi \;,\;y=\xi r\sin \phi ,  \label{dimpolar}
\end{equation}
and taking $\varepsilon =\xi /x_{0}$, we can express $z$ as we approach the
vortex as 
\begin{eqnarray}
z\left( 1+\varepsilon r\cos \phi ,\varepsilon r\sin \phi \right)  &=&1+\frac{%
\varepsilon ^{2}}{2}\frac{r^{2}}{1+\varepsilon r\cos \phi }  \nonumber \\
&\equiv &1+\frac{\varepsilon ^{2}\Delta ^{2}}{2}.  \label{zexpand}
\end{eqnarray}
We can then use this expansion of $z$ in the integral representation \cite
{Guns3} of the Legendre function 
\begin{equation}
Q_{\nu }(z)=\int_{0}^{\infty }\frac{ds}{\left( z+\sqrt{z^{2}-1}\cosh
s\right) ^{\nu +1}}  \label{qintrep}
\end{equation}
to obtain 
\begin{equation}
Q_{\frac{1}{2}}(1+\varepsilon ^{2}\Delta ^{2}/2)=\mathcal{O}(\varepsilon
^{2})+\int_{0}^{\infty }\frac{ds}{\left( 1+\varepsilon \Delta \cosh s\right)
^{3/2}}  \label{Qsmallr}
\end{equation}
Then we can note that 
\begin{eqnarray}
&&\int_{0}^{\infty }\frac{ds}{\left( 1+\varepsilon \Delta \cosh s\right)
^{3/2}}  \nonumber \\
&=&\int_{0}^{-\frac{\ln \varepsilon \Delta }{2}}\frac{ds}{\left(
1+\varepsilon \Delta \cosh s\right) ^{3/2}}  \nonumber \\
&&+\int_{-\frac{\ln \varepsilon \Delta }{2}}^{\infty }\frac{ds}{\left(
1+e^{s+\ln \frac{\varepsilon \Delta }{2}}+e^{-s+\ln \frac{\varepsilon \Delta 
}{2}}\right) ^{3/2}}  \label{integralsss}
\end{eqnarray}
and evaluate each of the two integrals using different expansions in $%
\varepsilon $ (thus evaluating (\ref{qintrep}) using a kind of minature
boundary layer theory of its own!). \ That is, in the first integral of the
RHS of (\ref{integralsss}) we simply expand in $\varepsilon \Delta $ and
integrate term by term; and in the second we use 
\begin{eqnarray}
&&\int_{-\frac{\ln \varepsilon \Delta }{2}}^{\infty }\frac{ds}{\left(
1+e^{s+\ln \frac{\varepsilon \Delta }{2}}+e^{-s+\ln \frac{\varepsilon \Delta 
}{2}}\right) }  \nonumber \\
&\doteq &\int_{\ln \frac{\sqrt{\varepsilon \Delta }}{2}}^{\infty }\frac{ds}{%
\left( 1+e^{s}\right) ^{3/2}}  \nonumber \\
&=&\left[ \frac{2}{\sqrt{1+e^{s}}}+\ln \frac{\sqrt{1+e^{s}}-1}{\sqrt{1+e^{s}}%
+1}\right] _{\ln \frac{\sqrt{\varepsilon \Delta }}{2}}^{\infty }
\end{eqnarray}
where in the second line, as hereafter, $\doteq $ means dropping a term of
order $\varepsilon ^{2}$.\newline
\newline
\ Combining both results we obtain 
\begin{eqnarray}
Q_{\frac{1}{2}}(1+\varepsilon ^{2}\Delta ^{2}/2) &\doteq &-\left( 2+\ln 
\frac{\varepsilon \Delta }{8}\right)   \nonumber \\
&\doteq &-2-\ln \frac{\varepsilon r}{8}+\frac{\varepsilon r\cos \phi }{2},
\end{eqnarray}
and hence 
\begin{eqnarray}
&&\!F(x_{0}+\xi r\cos \phi ,\xi r\sin \phi )  \nonumber \\
&=&\left( 1+\frac{\varepsilon r\cos \phi }{2}\right) \ln \frac{\varepsilon r%
}{8}  \nonumber \\
&&+2+\frac{\varepsilon }{2}r\cos \phi +\mathcal{O}(\varepsilon ^{2}).
\end{eqnarray}
\noindent \noindent 

Translating from the dual potential $F$ back to $\vec{A}$ according to (\ref
{Fdef}), to first order in $\varepsilon $ we therefore have 
\begin{eqnarray}
&&\lim_{\vec{x}\rightarrow \vec{x}_{0}}A_{r}\doteq \frac{1}{\xi }\left[ 
\frac{1}{r}+\frac{\varepsilon \cos \phi }{2}\ln \frac{\varepsilon r}{8}%
\right]  \\
&&\lim_{\vec{x}\rightarrow \vec{x}_{0}}A_{\phi }\doteq -\frac{1}{\xi }\left[ 
\frac{\varepsilon }{2}\left( 1+\ln \frac{\varepsilon r}{8}\right) \right]
\sin \phi .
\end{eqnarray}
Applying (\ref{dualA}) we then see 
\begin{eqnarray*}
\lim_{\vec{x}\rightarrow \vec{x}_{0}}\partial _{\phi }\theta  &\doteq &1+%
\frac{\varepsilon r\cos \phi }{2}\ln \frac{\varepsilon r}{8} \\
\lim_{\vec{x}\rightarrow \vec{x}_{0}}\partial _{r}\theta  &\doteq &\frac{%
\varepsilon }{2}\left( 1+\ln \frac{\varepsilon r}{8}\right) \sin \phi 
\end{eqnarray*}
from which we can obviously extract the condensate phase field near the
vortex, 
\begin{equation}
\theta (r,\phi )=\phi +\varepsilon r\frac{\sin \phi }{2}\ln \frac{%
\varepsilon r}{8}+\mathcal{O}(\varepsilon ^{2}).  \label{F2a}
\end{equation}
Note that there is no trace of Euler's constant in this equation.

\section{Inner solution: Gross-Pitaevskii}

\subsection{The perturbative problem}

In the inner zone we work entirely in terms of the dimensionless polar
co-ordinates centered on the vortex: $x=x_{0}+\xi r\cos \phi $, $y=\xi r\sin
\phi $. \ (Since the actual condensate density at the vortex center
vanishes, it is perhaps worth clarifying that for the local healing length
`at' the vortex $\xi =\left( g\rho _{0}\right) ^{-1/2}$ we use for $\rho _{0}
$ the background Thomas-Fermi density extrapolated to the vortex location.)
\ To obtain our inner region solution, we expand the potential to first
order about the vortex position: $V(\vec{x})=\lambda x=\lambda x_{0}\left[
1+\varepsilon r\cos \phi +\mathcal{O}(\varepsilon ^{2})\right] $. It will
then be convenient to rescale the wave function, defining 
\begin{equation}
\Psi =e^{-i\mu t/\hbar }\sqrt{\rho _{0}}\psi \left( \vec{x}-\varepsilon \vec{%
\beta}ct\right) 
\end{equation}
where $c=\hbar /(M\xi )$ is the local speed of sound at the vortex, and$\ 
\vec{v}_{vtx}=\varepsilon \vec{\beta}c$ is the vortex velocity, as yet
unknown. (We have reduced the number of symbols to be defined by
anticipating the fact that the vortex velocity will be order $\varepsilon $%
.) \ Then writing $\psi =\psi _{0}+\varepsilon \psi _{1}+...$ ,\ we find 
\begin{equation}
-\frac{1}{2}\left[ \partial _{r}^{2}+\frac{1}{r}\partial _{r}+\frac{1}{r^{2}}%
\partial _{\phi }^{2}\right] \psi _{0}+\left[ |\psi _{0}|^{2}-1\right] \psi
_{0}=0  \label{psi0}
\end{equation}
as our zeroth order equation. \ The only solution which corresponds to a
singly quantized vortex at $r=0$ is $\psi _{0}=f(r)e^{i\phi }$ where $f(r)$
can be taken as real, and satisfies  \  
\begin{equation}
\frac{1}{2}\left[ \partial _{r}^{2}+\frac{1}{r}\partial _{r}\right] f=\frac{f%
}{2r^{2}}+\left[ f^{2}-1\right] f  \label{vortf}
\end{equation}
and $f(r\rightarrow \infty )=1$. \ Note that a vortex velocity of order $%
\varepsilon ^{0}$ would require a $\vec{v}_{vtx}\cdot \vec{\nabla}\psi _{0}$
term on the RHS of (\ref{psi0}), and the only well-behaved vortex solution
with such a term present is $e^{iM\vec{v}\cdot \vec{x}/\hbar }\psi _{0}$;
but this would imply a phase gradient of order $\varepsilon ^{0}$ as one
approaches the outer zone ($r\rightarrow R/\xi $), and this would be
inconsistent with our outer solution (\ref{F2a}).

At first order, though, we find 
\begin{eqnarray}
&&-\frac{1}{2}\left[ \partial _{r}^{2}+\frac{1}{r}\partial _{r}+\frac{1}{%
r^{2}}\partial _{\phi }^{2}\right] \psi _{1}+\left[ 2f^{2}-1\right] \psi
_{1}+\psi _{0}^{2}\psi _{1}^{\ast }  \nonumber \\
&=&r\cos \phi \psi _{0}-i\left( \beta _{x}\cos \phi +\beta _{y}\sin \phi
\right) \partial _{r}\psi _{0}  \nonumber \\
&&+\left( \beta _{y}\cos \phi -\beta _{x}\sin \phi \right) \frac{\psi _{0}}{r%
}\;.  \label{psi1}
\end{eqnarray}
It is not easy to solve this equation for $\psi _{1}$; but we are actually
only interested in two pieces of information. \ We need to know the
asymptotic behaviour of $\psi _{1}$ many healing lengths away from the
vortex center, so that we can smoothly match the inner solution to the
outer. \ And we need to determine $\vec{\beta}$, which will be fixed by the
requirement that $\psi _{1}$\ does not blow up as $r$ increases. \ Following
Rubinstein and Pismen \cite{RP}, we will be able to obtain this information
without solving (\ref{psi1}) explicitly. \ 

At large $r$ we can write $\psi _{1}\rightarrow e^{i\phi }[S+iT]$ and find $%
S\rightarrow \frac{r}{2}\cos \phi .$ \ Then since $f\rightarrow 1-(2r)^{-2}$
at large $r$, the leading terms in $T$ are driven by the $-ir^{-2}\partial
_{\phi }S$ crossterm, giving the asymptotic equation $\nabla
^{2}T=r^{-1}\sin \phi $, with the solution 
\begin{equation}
T\rightarrow r\left( \alpha _{x}\cos \phi +\alpha _{y}\sin \phi \right) +%
\frac{r}{2}\ln r\sin \phi \ 
\end{equation}
for coefficients $\vec{\alpha}$ that will be fixed by matching with the
outer zone. \ (We drop a constant which can obviously be absorbed in $\psi
_{0}$, and which is of no consequence anyway.) \ So we have 
\begin{eqnarray}
\lim_{r\rightarrow \infty }\Psi  &=&e^{i\phi }\sqrt{\rho _{0}}[1+\varepsilon 
\frac{r}{2}(\cos \phi +i\ln r\sin \phi )  \nonumber \\
&&\qquad \qquad \qquad \qquad +i\varepsilon \vec{\alpha}\cdot \vec{r}] 
\nonumber \\
&=&\sqrt{\rho }\exp i[\phi +\left( \varepsilon r/2\right) \left( \ln
r+2\alpha _{y}\right) \sin \phi   \nonumber \\
&&\qquad \qquad +\varepsilon r\alpha _{x}\cos \phi ]  \nonumber \\
&=&\sqrt{\rho }\exp i\left( \phi +\varepsilon \frac{r}{2}\ln \frac{%
\varepsilon r}{8}\sin \phi \right)   \label{damnyouscientificworkplace}
\end{eqnarray}
dropping terms of $\mathcal{O}(\varepsilon ^{2})$. \ In the last line we
have imposed matching with the outer solution (\ref{F2a}) to fix 
\begin{equation}
\alpha _{x}=0\quad ,\quad \alpha _{y}=\frac{1}{2}\ln \frac{\varepsilon }{8}.
\label{alphares}
\end{equation}

To constrain $\vec{\beta}$, note that differentiating (\ref{psi0}) with
respect to $x$ or $y\ $shows that $\partial _{x}\psi _{0}$ and $\partial
_{y}\psi _{0}$\ are two independent solutions to the homogeneous equation
for $\psi _{1}$. \ Writing $\mathcal{E}$ as an abbreviation for the
left-hand side of (\ref{psi1}), and $\mathcal{J}$ as an abbreviation for the
right-hand side, we integrate both sides of (\ref{psi1}) with $\partial
_{x}\psi _{0}^{\ast }$ out to the large dimensionless radius $R/\xi $. \
Integrating by parts and using our results for $\psi _{1}$ at large radius
reveals 
\begin{eqnarray}
&&\func{Re}\left[ \int_{0}^{\frac{R}{\xi }}rdr\oint d\phi \left[ e^{-i\phi
}\left( f^{\prime }\cos \phi +i\frac{f}{r}\sin \phi \right) \mathcal{E}%
\right] \right]   \nonumber \\
&=&\frac{1}{2}\func{Im}\oint d\phi \left[ e^{-i\phi }\sin \phi \left(
\partial _{r}\psi _{1}+r^{-1}\psi _{1}\right) \right] _{r=R/\xi }  \nonumber
\\
&=&\frac{\pi }{2}\left[ \ln \frac{\varepsilon R}{8\xi }+\frac{1}{2}\right] =%
\frac{\pi }{2}\left[ \ln \frac{R}{\xi }+\frac{1}{2}+2\alpha _{y}\right]  
\nonumber \\
&=&\func{Re}\left[ \int_{0}^{\frac{R}{\xi }}rdr\oint d\phi \left[ e^{-i\phi
}\left( f^{\prime }\cos \phi +i\frac{f}{r}\sin \phi \right) \mathcal{J}%
\right] \right]   \nonumber \\
&=&\pi \int_{0}^{\frac{R}{\xi }}drff^{\prime }\left[ r^{2}+2\beta _{y}\right]
\nonumber \\
&=&\pi \left[ \beta _{y}+\frac{R^{2}}{2\xi ^{2}}(1-\frac{\xi ^{2}}{2R^{2}}%
)-\int_{0}^{\frac{R}{\xi }}dr\,rf^{2}\right] \;.
\end{eqnarray}
This allows us to obtain 
\begin{eqnarray}
\beta _{y}-\alpha _{y} &=&\lim_{R/\xi \rightarrow \infty }\left[ \frac{1+\ln
\left( R/\xi \right) }{2}-\int_{0}^{\frac{R}{\xi }}dr\,r(1-f^{2})\right]  
\nonumber \\
&=&\frac{1}{2}\left[ 1+\ln \frac{R}{\xi }+\int_{0}^{\frac{R}{\xi }}dr\,\left[
r\frac{f^{\prime \prime }}{f}+\frac{f^{\prime }}{f}-\frac{1}{r}\right] %
\right]   \nonumber \\
&\doteq &\frac{1}{2}\left[ 1-\ln f^{\prime }(0)+\int_{0}^{\infty }dr\,r\frac{%
f^{\prime \prime }}{f}\right]   \nonumber \\
&\doteq &-0.114\;.  \label{b1y}
\end{eqnarray}
In the second line we have used (\ref{vortf}) to re-write $(1-f^{2})$ in the
integrand, and in the last line\ the numerical evaluation comes from a
numerical solution for $f(r)$ . \ Similarly, using the $\partial _{y}\psi
_{0}=e^{i\phi }\left[ f^{\prime }\sin \phi +ir^{-1}f\cos \phi \right] $
solution, we find 
\begin{equation}
\beta _{x}=0.  \label{b1x}
\end{equation}

(The numerical result in (\ref{b1y}) was obtained with \textit{Mathematica} 
\cite{Math},\textit{\ }using two different methods of numerical solution for 
$f(r)$ (shooting and relaxation), whose results agreed with each other.
Quite stringent settings of the options for starting step sizes, etc., were
required to obtain this agreement, and in both methods the singularities at $%
r=0$ had to be regulated, for example by replacing $r\rightarrow \sqrt{%
r^{2}+10^{-20}}$ in the singular co-efficients in (\ref{vortf}).\ Our result
does not quite agree with the evaluation reported in \cite{RP}: their value
of $0.405$ for their quantity $\ln a_{1}$ corresponds to a value of $-0.126$
for our quantity $\beta _{y}-\alpha _{y}$.) \ 

Combining (\ref{alphares}) and (\ref{b1y}), we obtain the total vortex
velocity. \ It is in the negative $y$ direction: parallel to the TF surface,
and in the direction of the fluid flow between the vortex and the surface. \
Extracting explicitly the $x_{0}$ dependence hidden in $\varepsilon =\sqrt{%
2\delta ^{3}/x_{0}^{3}}$, we can express the magnitude $v_{vtx}$\ of the
vortex velocity purely in terms of $x_{0}$ and surface parameters: 
\begin{eqnarray}
v_{vtx} &=&\varepsilon c\left| \beta \right| =\frac{\xi }{x_{0}}\frac{\hbar 
}{M\xi }\left| \beta \right|   \nonumber \\
&=&\frac{\hbar }{Mx_{0}}\left[ \frac{1}{4}\ln \left( \frac{32x_{0}^{3}}{%
\delta ^{3}}\right) +0.114\right]   \nonumber \\
&=&v_{c}\frac{\delta }{x_{0}}\left[ \frac{3}{4}\ln \left( \frac{x_{0}}{%
\delta }\right) +0.980\right] \;  \label{main}
\end{eqnarray}
where 
\begin{equation}
v_{c}=\frac{\hbar }{M\delta }
\end{equation}
is the surface characteristic (and critical \cite{me})\ velocity. \ Eqn. (%
\ref{main}) is the first main result of this paper; it is illustrated
graphically by Figure 3. \ It will be accurate for vortices in
quasi-two-dimensional condensates as long as the vortex distance from the TF
surface $x_{0}$ is much larger than the surface depth $\delta $ but much
smaller than the TF radius. \ For quasi-2D condensates of size comparable to
current three-dimensional condensates, this will be a significant regime of
validity. \ Of course, even apart from its realism, (\ref{main}) remains
instructive as an accurate result for an idealized problem. 
\begin{figure}[tbp]
\includegraphics[width=.475\textwidth]{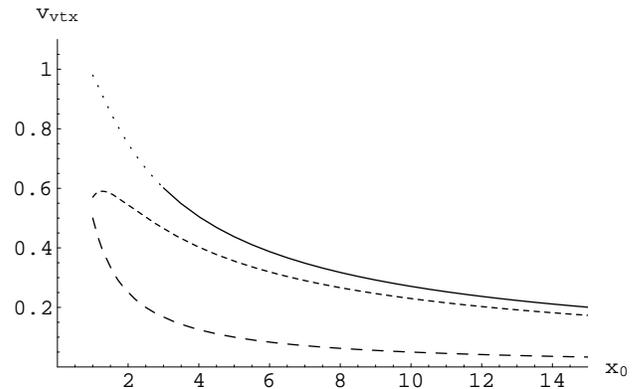}
\caption{Velocity of a vortex parallel to a Thomas-Fermi surface in a
quasi-two-dimensional condensate (upper curve). \ The vertical axis is
velocity in units of the surface mode critical velocity $\hbar /(M\protect%
\delta )$; the horizontal axis is distance of the vortex center from the TF
surface, in units of the surface depth $\protect\delta $. \ As a leading
order result in $\protect\delta /x_{0},$ the curve is not meaningful to the
left $x_{0}=\protect\delta $. \ \ The dashed curves, presented for
comparison, are the Rubinstein-Pismen result (short dashes), and the
velocity $\hbar /(2Mx_{0})$ of a vortex the same distance from a hard wall
in a condensate of constant density (long dashes). The sum of the two dashed
curves is quite a good approximation to the solid curve. \ }
\end{figure}

\subsection{Free energy and vortex penetration}

If dissipation occurs in a frame moving with respect to the condensate with
velocity $v_{dis}$, then the vortex will drift towards or away from the
surface, in order to minimize the free energy 
\begin{equation}
G=\frac{\pi \hbar ^{2}}{2Mg\delta ^{2}}\left( E-p\frac{v_{dis}}{v_{c}}%
\right) ,
\end{equation}
where the prefactor is a surface energy scale, so that the dimensionless
energy $E$ and $y$ component of momentum $p$ are given by 
\begin{eqnarray}
E &=&\frac{x_{0}}{\pi \rho _{0}\delta }\int d^{2}x\,\left[ \frac{1}{2}\left| 
\vec{\nabla}\Psi \right| ^{2}+\frac{g}{2}\left( \left| \Psi \right| ^{2}-%
\frac{Fx}{g}\right) ^{2}\right]  \\
p &=&\frac{x_{0}}{\pi \rho _{0}\delta ^{2}}\func{Im}\int dxdy\;\Psi ^{\ast
}\partial _{y}\Psi .
\end{eqnarray}
In this expression for $E$ we have subtracted off the Thomas-Fermi free
energy of the vortex-free condensate. \ We will now evaluate this free
energy to leading order in $\varepsilon $ using our results from above. \ As
above, we will compute the inner and outer zone contributions separately,
and add them. \ In this sum, all dependences on the inner zone size $R\sim
\varepsilon ^{-1/2}\xi $ necessarily cancel out.

\subsubsection{Computing $E$}

Since we know that $\nu _{vtx}$ is of order $\varepsilon $, it is not hard
to see that the order $\varepsilon $ terms in the inner component $E_{in}$
of $E$ vanish, and up to corrections of order $\varepsilon ^{2}$ we have 
\begin{eqnarray}
E_{in} &=&\frac{x_{0}}{\delta }\int_{0}^{\frac{R}{\xi }}dr\;r\left[
f^{\prime 2}+\frac{f^{2}}{r^{2}}+\left( f^{2}-1\right) ^{2}\right]  
\nonumber \\
&=&\frac{x_{0}}{\delta }\int_{0}^{\frac{R}{\xi }}\!\!dr\;\partial _{r}\left[
rff^{\prime }+r^{2}f^{2}-\frac{r^{2}f^{4}+f^{2}-r^{2}f^{\prime 2}}{2}\right] 
\nonumber \\
&&+\frac{x_{0}}{\delta }\int_{0}^{\frac{R}{\xi }}dr\;r\left( 1-2f^{2}\right) 
\nonumber \\
&=&\frac{x_{0}}{\delta }\left[ \ln \frac{R}{\xi }-2\left( \beta
_{y}-a_{y}\right) +\frac{1}{2}\right] 
\end{eqnarray}
where in the last line we also drop terms of order $\left( \xi /R\right) ^{2}
$, which are not only of order $\varepsilon $, but will also be cancelled by
terms from the outer component $E_{out}$.

In the outer zone we must integrate the energy density over the entire half
plane, except for the inner zone circle of radius $R\sim \varepsilon
^{1/2}x_{0}$. \ Denoting integration over this region with the subscript $%
\mathcal{A}$ we have, to leading order in $\varepsilon $ 
\begin{eqnarray}
E_{out} &=&\frac{x_{0}}{\pi \delta }\int\limits_{\mathcal{A}}dxdy\;\frac{%
\rho }{\rho _{0}}\left| \vec{\nabla}\theta \right| ^{2}=\frac{x_{0}}{\delta }%
\int\limits_{\mathcal{A}}dxdy\;\frac{x_{0}}{x}\left| \vec{\nabla}F\right|
^{2}  \nonumber \\
&=&\frac{x_{0}^{2}}{\pi \delta }\int\limits_{\mathcal{A}}dxdy\;\vec{\nabla}%
\cdot \left( \frac{F}{x}\vec{\nabla}F\right) =\frac{x_{0}^{2}}{\pi \delta }%
\oint\limits_{\partial \mathcal{A}}d\vec{S}\cdot \frac{F}{x}\vec{\nabla}F 
\nonumber \\
&=&-\frac{x_{0}}{\pi \delta }\oint\limits_{r=R/\xi }d\phi \;F\partial _{r}F 
\nonumber \\
&=&-\frac{x_{0}}{\delta }\left( \ln \frac{\varepsilon R}{8\xi }+2\right) .
\end{eqnarray}
To obtain the second line from the first we use (\ref{hydroedge}), where the
RHS\ vanishes everywhere in $\mathcal{A}$. \ Within the second line we use
the divergence theorem, where the surface terms at $x=0$ and at infinity all
vanish, so that the only contribution is on the boundary with the inner zone
at the circle of radius $R$ about $x_{0}.$ \ So putting $E_{in}$ and $E_{out}
$ together we have 
\begin{eqnarray}
E &=&-\frac{x_{0}}{\delta }\left( \frac{3}{2}+2\beta _{y}\right)   \nonumber
\\
&=&2\frac{x_{0}}{\delta }\left[ \frac{3}{4}\left( \ln \frac{x_{0}}{\delta }%
-1\right) +0.980\right] .
\end{eqnarray}
Writing $E$ as we have in the last line emphasizes that $dE/dx_{0}$ is
proportional to $v_{vtx}$, indicating that $x_{0}$ and $v_{vtx}$ are
canonically conjugate, as one expects for a vortex. \ 

\subsubsection{Computing $p$}

To compute $p$ it is very helpful to note that the dual potential $F$ can be
extended into the inner zone, by using the continuity equation and the fact
that the density is constant in the frame co-moving with the vortex to obtain
\begin{eqnarray}
0 &=&\left( M/\hbar \right) \left[ v_{vtx}\partial _{y}\rho -\dot{\rho}%
\right]   \nonumber \\
&=&\vec{\nabla}\cdot \left[ \rho \left( \vec{\nabla}\theta +\frac{M}{\hbar }%
\hat{y}v_{vtx}\right) \right] 
\end{eqnarray}
which allows us to define
\begin{eqnarray}
\rho \partial _{y}\theta  &=&\rho _{0}\partial _{x}F+\frac{M}{\hbar }%
v_{vtx}\left( \rho -\frac{x}{x_{0}}\rho _{0}\right)  \\
\rho \partial _{x}\theta  &=&-\rho _{0}\partial _{y}F\;.
\end{eqnarray}
Taking this definition of $F$ into the outer zone, it clearly agrees with
our previous one up to post-hydrodynamic corrections. \ (Alternatively one
could repeat the outer zone analysis using this slightly different
definition of $F$, and confirm that no differences arose up to order $%
\varepsilon ^{2}$.) \ Furthermore one can show from the Gross-Pitaevskii
equation that this extended $F$ is perfectly regular as $r\rightarrow 0$,
where it behaves as $\int dr\,r^{-1}f^{2}$. \ We can therefore write 
\begin{eqnarray}
p &=&\frac{x_{0}}{\pi \delta ^{2}}\int dxdy\;\frac{\rho }{\rho _{0}}\partial
_{y}\theta   \nonumber \\
&=&\frac{x_{0}}{\pi \delta }\frac{v_{vtx}}{v_{c}}\int dxdy\left( \frac{\rho 
}{\rho _{0}}-\frac{x}{x_{0}}\right) \;  \label{psurf} \\
&&-\frac{x_{0}}{\pi \delta ^{2}}\int_{-\infty }^{\infty }dy\left[
F(0,y)-\lim_{x\rightarrow \infty }\left[ F(x,y)\right] \right] .  \nonumber
\end{eqnarray}
The leading contribution to $p$ is the last line of (\ref{psurf}). \ Its
evaluation is very simply obtained from (\ref{as2}): 
\begin{eqnarray}
p &\doteq &\frac{x_{0}}{\pi \delta ^{2}}\lim_{x\rightarrow \infty
}\int_{-\infty }^{\infty }dy\,F(x,y)\,  \nonumber \\
&=&-\frac{x_{0}^{2}}{2\delta ^{2}}\lim_{x\rightarrow \infty }\int_{-\infty
}^{\infty }\frac{dy}{x}\left( 1+\frac{y^{2}}{x^{2}}\right) ^{-3/2}  \nonumber
\\
&=&-\frac{x_{0}^{2}}{2\delta ^{2}}\int_{-\infty }^{\infty }\frac{dy}{\left(
1+y^{2}\right) ^{3/2}}=-\frac{x_{0}^{2}}{\delta ^{2}}.
\end{eqnarray}
This leaves out the density deficit integral in the second line of (\ref
{psurf}), which can be shown to be of order $\delta /x_{0}$, and hence
smaller than the leading term by a factor of $\varepsilon ^{2}$.

\subsubsection{Free energy curve}

We therefore have, up to corrections of order $\varepsilon ^{2},$ 
\begin{equation}
G=\frac{\pi \hbar ^{2}}{2Mg\delta ^{2}}\left[ \frac{x_{0}}{\delta }\left(
0.46+\frac{3}{2}\ln \frac{x_{0}}{\delta }\right) +\frac{v_{dis}}{v_{c}}%
\left( \frac{x_{0}}{\delta }\right) ^{2}\right]   \label{edgefree}
\end{equation}
which is plotted in Figure 4. \ At least within the hydrodynamic
approximation, the energetic barrier to vortex penetration has clearly
disappeared for $|v_{dis}|>v_{c}$. \ This is the second main result of this
paper. \ 
\begin{figure}[tbp]
\includegraphics[width=.475\textwidth]{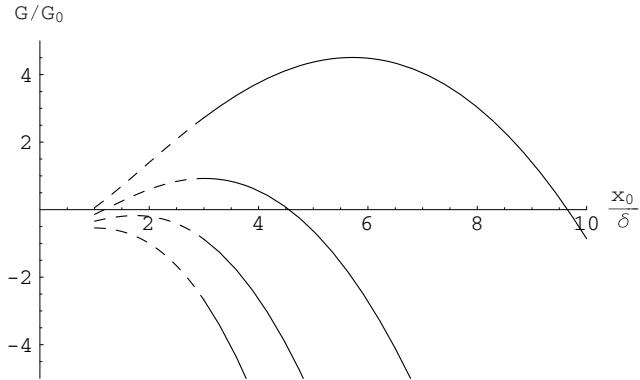}
\caption{Free energy $G/G_{0}=E-pv_{dis}/v_{c}$, where $G_{0}=\protect\pi %
\hbar ^{2}/(2Mg\protect\delta ^{2})$, $\protect\delta =\left( 2M\protect%
\lambda /\hbar ^{2}\right) ^{-1/3}$, and $v_{c}=\hbar /(M\protect\delta )$.
\ Horizontal axis is $x_{0}/\protect\delta $, distance of vortex center from
Thomas-Fermi surface in units of surface depth $\protect\delta $. \ From
uppermost to lowermost, the curves are for $v_{dis}/v_{c}=-0.4,-0.6,-0.8,-1.$
\ The plots are shown dashed for $x_{0}/\protect\delta <3$, and stop at $%
x_{0}/\protect\delta =1$, to reflect that the curves are leading order
results in $\protect\delta /x_{0}$. \ \ }
\label{alpha}
\end{figure}
Since we are considering a vortex with counter-clockwise circulation,
located along the positive $x$ axis away from a TF surface on the $y$ axis,
it is to be expected that a negative $v_{dis}$ makes vortex penetration
become energetically favourable. \ 

\subsubsection{Application to rotating harmonic traps}

For a harmonically trapped condensate of spatial size $R_{TF}$, the
nonlinearity of the trapping potential near the TF surface, and the
curvature of the TF surface, can both be neglected in the limit where $%
x_{0}/R_{TF}\rightarrow 0$. \ Hence our results can be applied to
harmonically trapped condensates if we regard them as leading order
approximations in both $\delta /x_{0}$ and $x_{0}/R_{TF}$. \ If we consider
a rotational symmetric trap, we take $R_{TF}$ to be the Thomas-Fermi radius
\ in the plane of symmetry, and then assume either a quasi-two-dimensional
`pancake' trap, or a quasi-cylindrical `extreme cigar' trap in which vortex
lines are parallel to the long axis $z$. \ In the latter case the $z$ length
of the condensate will only appear as an overall factor in the free energy,
and so effectively these two cases are exactly similar. \ We can translate
our parameters into those commonly used for harmonic trapped condensates by
noting that the potential gradient at the TF surface is just 
\begin{equation}
\lambda =M\omega ^{2}R_{TF}
\end{equation}
where $\omega $ is the radial trap frequency. \ This yields 
\begin{equation}
\delta =\left( \frac{\hbar ^{2}}{2M^{2}\omega ^{2}R_{TF}}\right) ^{1/3}
\end{equation}
and 
\begin{equation}
\varepsilon =\frac{\hbar }{M\omega \sqrt{x_{0}^{3}R_{TF}}}=\frac{a_{0}^{2}}{%
x_{0}^{3/2}R_{TF}^{1/2}}
\end{equation}
for the trap size $a_{0}=\left( M\omega /\hbar \right) ^{-1/2}$. \ If we
estimate the size of $\varepsilon $ in an effectively 2D condensate by
supposing dimensions typical of current large three-dimensional condensates,
taking $a_{0}\sim \mu $m and $R_{TF}\sim 25\mu $m gives $\varepsilon
\lesssim 0.2$ for $x_{0}\gtrsim a_{0}$. \ So we see that the calculation
should be accurate for vortices a few microns inside the TF surface. \ (It
should be emphasized once again that three dimensional effects such as
bending of vortex lines are known to be present in current experiments, but
are entirely absent in our model. \ It seems reasonable to hope that vortex
bending will be slight and have small impact in highly prolate or highly
oblate traps, which should approach either our 2D or cylindrical limits; but
it is difficult to extend the present analysis far enough to support this
hope with calculations. \ The prospect that an initially straight vortex
line might bend more and more as it moved seems difficult to rule out.)

We can also obviously interpret $v_{dis}=\Omega R_{TF}$ where $\Omega $ is
the frequency at which a perturbation to the potential is rotated in order
to stir the condensate. \ With this translation of $v_{dis}$, we can compare
Figure 4 to the right-hand edge of Figure 5 from Reference \cite{SF2}, and
compare (\ref{edgefree}) to Eqn. (49) of \cite{SF2} (`SF49') in the limit
where $\zeta _{0}\rightarrow 1-2x_{0}/R$ in that equation. \ An essentially
similar equation is presented in \cite{KPSZ} as their Eqn. (9). \ (Both
these works examine three-dimensionally harmonic traps, in which, however,
vortex curvature is neglected, so that the length of the vortex line is
proportional to $x_{0}^{1/2}$. \ Hence in their cases energy and momentum
scale as $x_{0}^{3/2}$ and $x_{0}^{5/2}$, respectively, rather than $x_{0}$
and $x_{0}^{2}$ as in ours.) \ Equation SF49 was derived by assuming that,
for a rotationally symmetric trap, the phase field of the vortex would be
well approximated throughout the condensate by a simple $e^{i\phi }$ with $%
\phi $ the usual polar co-ordinate centered on the vortex. \ Unless the
vortex is very close to the center of the trap, this \textit{ansatz}
violates continuity significantly over most of the condensate. And we know
from our results above that it significantly exaggerates the degree to which
the velocity field of a vortex centered near the TF surface extends into the
bulk of the condensate. \ Hence according to SF49, the rotation frequency $%
\Omega _{\nu }$ \cite{FCS}\ at which the free energy becomes negative for
all $x_{0}>\delta $ is an overestimate (\textit{i.e.} is greater than $%
v_{c}/R_{TF}$) by the significant factor $(5/4)\ln (R_{TF}/\xi _{0})$. \ So
while discussions of vortex free energies based on the simple $e^{i\phi }$
ansatz will typically be qualitatively sound, they can easily be inaccurate
by factors of two or more, and cannot be used to obtain precise predictions
for critical velocities or stirring frequencies.

\section{Discussion}

\subsection{Magnus effect and infrared dressing}

The vortex velocity component $\beta _{y}-\alpha _{y}$ is perhaps the least
trivial aspect of the vortex motion, inasmuch as it describes a component of
the motion that is determined by the inner zone analysis alone, independent
of the background fluid velocity $\alpha _{y}$ in its immediate
neighbourhood. (The latter is nontrivial to determine, from the outer zone
hydrodynamics, but trivial in the way it moves the vortex). \ As first
pointed out by Rubinstein and Pismen \cite{RP}, this intrinsic motion is
along contours of constant Thomas-Fermi density.\ \ We can understand it
qualitatively by noting that the vortex is a bubble-like density defect, and
so experiences a buoancy force in the opposite direction to the trap force.
\ Due to the Magnus effect which dominates vortex motion in superfluids and
normal fluids alike, this force produces not an acceleration, but a velocity
at right angles: \ 
\begin{equation}
\vec{v}_{Mag}=-\frac{\vec{F}\times \hat{z}}{M\kappa \rho }=-\frac{\vec{F}%
\times \hat{z}}{2\pi \hbar \rho },
\end{equation}
since in our case the vortex circulation $\kappa $ is $2\pi \hbar /M$. \ (If
we had taken a vortex swirling in the opposite direction, we would indeed
have found $\beta _{y}$\ to have the opposite sign.) \ 

The precise magnitude of the buoancy force, and hence of the vortex
velocity, is nontrivial, because the naive buoancy force 
\begin{eqnarray}
\vec{F} &=&\vec{\nabla}V\int d^{2}x\left( \rho _{0}-|\Psi |^{2}\right)  
\nonumber \\
&=&-\hat{x}\varepsilon c\left( 2\pi \rho \hbar \right) \int dr\,r(1-f^{2})
\end{eqnarray}
is logarithmically divergent. \ What \cite{RP} showed first, and our own
inner zone analysis has reproduced, is that the buoancy force is
renormalized. \ The longer range effects of the potential gradient produce a
logarithmic distortion of the vortex flow pattern, and since the classic
Magnus effect is for a strictly cylindrical flow, this distortion produces
the counterterms $(1+\ln R/\xi )/2$ in the first line of (\ref{b1y}). \ 

On the other hand, we can see from (\ref{damnyouscientificworkplace}) that
the flow distortion in $\psi _{1}$ implies a velocity field component $%
\propto \hat{y}\ln r$, which is essentially indistinguishable in any range
of $r$\ from a background velocity perpendicular to the potential gradient.\
\ So the distinction between the intrinsic $\left( \beta _{y}-\alpha
_{y}\right) $\ and ambient $\alpha _{y}$ components of vortex motion is
somewhat artificial, and it is more natural to consider the whole flow
pattern as a `dressed vortex'.

Previous analytic studies of vortices in BECs have often neglected this
dressing effect, by assuming that the phase field $e^{i\theta (\vec{x})}$\
of a vortex always remains $e^{i\phi }$ in polar co-ordinates centred on the
vortex. \ Close to the core of a vortex this is indeed a good approximation,
as long as `close' means that the background condensate density has not
varied appreciably. \ If further away from the core the background density
varies (and is not rotationally symmetric about the vortex), then thia
simple \textit{ansatz} for the phase field obviously fails to satisfy the
continuity condition. \ Hence at distances from the vortex core that are on
the scale of the trapping potential's variation, the phase field will depart
significantly from $e^{i\phi }$. \ Furthermore, since the corrections to $%
e^{i\phi }$ are caused by density variations on the potential scale, their
own spatial scale will be of this same order. \ And this means that the
corrections will extend into the vortex core region, in the form a `local
ambient' flow.\ This flow is constant on the healing length scale, but it
exists just because the vortex is present in the inhomogeneous sample, and
in this sense can be considered part of the vortex: it is a component of the
vortex's infrared dressing. \ \ \ 

\subsection{Against image vortices}

The simplest example of this infrared dressing phenomenon is already well
known, as it occurs in the special case of inhomogeneity that is a `hard
wall' boundary on an otherwise homogeneous sample. \ Here the nontrivial
requirement of continuity far from the vortex is simply the constraint that
there be no flow through the hard wall surface. \ In this special case the
hydrodynamic problem to be solved for the vortex phase field is simply the
Laplace equation, with Dirichlet boundary conditions on the surface. \ For
many shapes of surface, the method of images is a technical trick that
produces the required solution. \ This convenient technique has perhaps had
the unfortunate side-effect of obscuring the general phenomenon of vortex
infrared dressing, by giving rise to an impression that `image vortices' are
the only significant effect of inhomogeneity. \ In fact the method of images
is restricted to conveniently symmetrical equations and boundary conditions,
and it requires an exact solution which is known in the Laplace case, but
not in general. \ It is of no help in the case of a Thomas-Fermi surface,
because the single vortex solution automatically satisfies the correct
boundary condition. \ And it is not even applicable for a hard wall surface,
unless the density profile within the wall has a fortunate form. \ (For
example, the solution obtained in this paper will allow image solutions for
a hard wall surface perpendicular to the gradient of a linear potential, but
not for walls at other angles.) \ 

Thinking about image vortices will typically allow a qualitative
understanding of how vortices will move near a surface; but this will not be
quantitatively reliable. \ For those strongly attached to the image vortex
picture, the singularity of $Q_{\nu }(z)$ at $z=-1,$ corresponding to $%
x=-x_{0}$, may reveal an image vortex even in the case solved above. \ But
since image vortices are actually just a mathematical trick, for
representing the physical effect of a surface, it would arguably be just as
well not to rely on them as a conceptual tool beyond their regime of real
applicability. \ Instead of picturing repulsive or attractive image
vortices, it is not so hard just to think about the accelerated flow through
the `channel' between a vortex and a surface, and so obtain a qualitative
understanding that is equally convenient and more genuine. \ 

On the other hand, using the Laplace image vortex velocity $\hbar /(2Mx_{0})$
to estimate the size of surface effects on vortices is obviously better than
nothing, if the full hydrodynamic problem is intractable (as it may well
be). \ Combining this crude image vortex theory with the local theory of
Rubinstein and Pismen \cite{RP} does seem to work surprisingly well for the
particular case analysed in this paper; but there are no grounds for
expecting it to be generally accurate. \ Using such uncontrolled
approximations to guide experimental design might be reasonable, given the
many unknown factors present in the early stages of an experiment. \ But
even fairly substantial disagreements between experimental measurements, and
theoretical predictions based on such zeroth-order estimates, would be no
evidence of novel phenomena.

\subsection{Estimating the critical velocity for vortex `nucleation'}

Since the vortex flow field in a homogeneous sample is already long ranged,
it is possible that the infrared dressing which occurs in an inhomogeneous
background may distort the long ranged velocity field by actually reducing
its extent. \ We have seen in this paper that for a vortex near a
Thomas-Fermi surface this sort of screening effect does indeed occur, so
that the vortex is a well localized structure. \ As we discussed briefly at
the end of the last Section, this effect can be a significant correction on
estimates of the velocity of stirring needed to drive vortices into a
condensate. \ 

Since surface Bogoliubov modes\ can be considered as the motion of vortices
in the ultra-dilute tail of the condensate density profile extending outside
the TF surface, the present paper can be regarded as a complement to \cite
{me}, showing how vortices may behave soon after they have `nucleated' (that
is, entered the condensate cloud). \ As we have mentioned, our hydrodynamic
and boundary layer approximations are only valid if the depth of the vortex
inside the TF surface is much greater than the local surface depth, and also
much smaller than the sample size. \ (Hence the applicability of the plane
linear model for vortices inside the condensate is somewhat more restricted
than for the Bogoliubov surface mode spectrum in \cite{me}, where it is only
required that the surface length be much less than the sample size.) \
Because of its inapplicability at small $x_{0}$, Figure 4 is not really
adequate to derive a precise result for the critical $v_{dis}$ above which
vortices will tend to enter the condensate spontaneously; but one can deduce
from it with good confidence that this critical value must be above $v_{c}/2$%
. \ The lowest curve of Figure 4 certainly shows that, above the surface
mode critical velocity $v_{c}$, no barrier to vortex penetration will remain
within the hydrodynamic region. \ This implies that the limitation on vortex
penetration is in the perturbative region, where the surface mode analysis
of \cite{me} fixes the critical velocity at $v_{c}=\hbar /(M\delta )$. \ And
this means that vortices enter a condensate through an ordinary instability,
and not by a quantum or thermal barrier-crossing process, which is usually
what is meant by the term `nucleation'. But two warnings must be attached to
this conclusion.

The first is that neither the surface mode analysis of \cite{me} nor the
boundary layer theory of the present paper are valid for vortices within a
few $\delta $ of the TF surface, and so one might in principle worry that a
narrow energetic barrier might still exist, within this region, at
velocities above $v_{c}$. \ Resolving this concern analytically would be
very difficult, but it can be dismissed with physical arguments. \ The TF
surface is not a physical surface, not a skin or a wall; it is a place where
the condensate density almost vanishes. \ In the neighbourhood of the TF
surface one can expect to see a transition between inner and outer regimes,
but there is simply not enough of anything there for this interface to
constitute a third regime of its own. \ Furthermore, vortices whose centers
are within a surface depth of the surface have core sizes on the order of $%
\delta $ as well, so that a vortex centered on the TF surface already
extends into the hydrodynamic regime on one side, and the perturbative
regime on the other. \ This makes it very implausible that a barrier arises
at the TF surface, when the free energy is monotonic in the same direction
on both sides of it. \ And, finally, experience with numerical integration
of dissipative 2D Gross-Pitaevskii equations has always shown that once
vortices begin sinking towards the TF surface, they pass through it and
enter the condensate without any difficulty. \ Hence we conclude that at
velocities above $v_{c}$ no barrier exists for vortices, at least in
condensates large enough for the linear density profile to be an accurate
local model.

The second warning is more important, which is that three dimensional
effects may perhaps hold vortex loops near the TF surface, simply because
for a vortex line to sink more deeply into a condensate it must typically
grow in length, adding the vortex `string tension' to the free energy. \ If
the vortex line bends and loops, this effect might be much greater than can
be accounted for by the factors of $x_{0}^{1/2}$ allowed in recent 3D
calculations assuming straight vortex lines. \ Just how strong such an
effect might be is far beyond the scope of this paper; but there have been
some indications in experiments that stirring a condensate at just above the
critical frequency generates twisting vortex loops that remain near the
edges of the cloud \cite{Jamil}.

\subsection{Summary and outlook}

This paper has presented an exactly solvable hydrodynamic problem, namely
the case of a point vortex in a plane linear background density profile,
which can supplement the hard wall as an example on which to base
understanding of general cases of vortices interacting with surfaces. \ This
particular solution is also directly relevant to the real problem of a
vortex near the Thomas-Fermi surface of a large condensate. \ It will even
be an accurate approximation in a realizable regime, because the velocity
field in this case actually falls off much more quickly than in the
homogeneous case, so that the longer-range effects of nonlinear potentials
and curved surfaces may be neglected more generally than one might initially
have expected. \ 

It is unfortunate, however, that the two conditions of a density gradient
and a surface are combined in this solvable case, because it therefore does
not afford us much intuition about how their effects may differ. \ It should
be clear from the preceding discussion that the effects found in this case
are not purely surface effects: if at some point the linear trapping
potential levelled off and then diminished, the TF surface would be replaced
by a mere local minimum in background density, but the density gradient
alone would still have effects on vortex dynamics. \ Vortices with density
inhomogeneity far from any surfaces should be analytically tractable in some
simple models, such as slightly varying periodic potentials. \ A few aptly
chosen numerical examples might be almost as instructive, and require less
effort.

\subsection{Acknowledgements}

The author thanks Eugene Zaremba for critical reading of an earlier version
of this paper. \ This work has been supported by the US\ National\ Science
Foundation through its grant for the Center for Ultracold Atoms. \

\end{document}